# Pt nanoparticles dispersed in a metal-organic framework as peroxidase mimics for colorimetric detection of GSH

Yanzheng Shu, Yanwei Chen, Guiye Shan

Centre for Advanced Optoelectronic Functional Materials Research, Key Laboratory for UV Light-Emitting Materials and Technology of the Ministry of Education, Northeast Normal University, Changchun, Jilin, 130024, China

* Corresponding author E-mail addresses: yanweichen@nenu.edu.cn (Yanwei Chen)

**ABSTRACT:**

Metal-organic skeleton materials have been widely used in catalysis with their porous structure and adsorption properties. Precious metal nanoparticles have good catalytic properties. If the noble metal nanoparticles are adsorbed on the MOFs surface, the active sites can be increased and the catalytic effect of the materials can be greatly improved. We successfully synthesized Pt@ZIF-8 in two steps, the average particle size of Pt nanoparticles is about 3 nm. Pt@ZIF-8 possesses peroxidase activity and can oxidize colorless TMB to oxTMB in the presence of hydrogen peroxide. The peroxide-like nature of Pt@ZIF-8 is consistent with Michaelis-Menten kinetics. Glutathione is a reducing substance that reduces blue oxTMB to colorless oxTMB. This colorimetric method achieves a simple, sensitive and intuitive detection of glutathione. The detection limit of this experiment is low, which is promising in biomolecular detection.

Translated with DeepL.com (free version)

## 1. Introduction

Nano-enzyme is a kind of nanomaterial with catalytic properties of enzymes [1]. Compared with natural enzymes [2], nano-enzyme has the advantages of high stability, low price, and reusable, which brings convenience to industrial production [3]. Noble metal nanomaterials have excellent physical, chemical properties and good biocompatibility, they are widely used in catalytic medical antibacterial hydrogen storage sensing and other aspects [4-7].

In recent years, many experiments have proved that loading precious metal nanoparticles onto other materials can enhance their stability [8-10]. For example, loading Au nanoparticles onto $WSe_2$ can greatly enhance the activity of the original nanomaterials [11], and dispersing Pt on the surface of $CuCo_2O_4$ like sea urchins can enhance the material's ability to decompose $H_2O_2$ and produce reactive oxygen species [12]. $Pt/WO_{2.72}$ has been reported to exhibit good peroxidase activity [13].

MOFs material is formed by the transition metal ion and organic ligand complex, metal part is usually called metal node, organic ligand is composed of oxygen atoms, carboxyl groups, carboxylic acids, etc. MOFs have unique advantages such as diverse structure, high specific surface area and adjustable porosity. Based on the above advantages, MOFs have been widely used in catalysis, gas adsorption, drug transportation and other fields [13-16].

Glutathione is a tripeptide composed of glutamic acid, cystine and glycine, which can promote the metabolic process of cells, and can also be used in the treatment of diseases, such as alcoholic liver disease, drug toxic liver disease, etc. Therefore, the detection of glutathione content is necessary. Compared with electrochemical detection [17], fluorescence detection [18], Raman detection [19] and other means, colorimetric detection [20-23] has the advantages of simple operation, low cost. The concentration range of the substance can be detected visually based on the change in colour of the chromogen.

In this work, we used ZIF-8 to load Pt nanoparticles. The Pt nanoparticles were in-situ grown on the surface of ZIF-8 to form Pt@ZIF-8. The structural properties and peroxidase activity of Pt nanoparticles were studied, and the results were used for colorimetric detection of glutathione (Scheme 1).

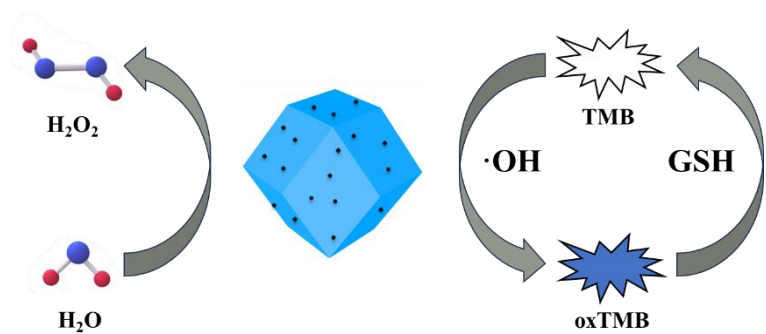

Scheme 1. Schematic graphic for the synthesis of Pt@ZIF-8 nanocrystals as the colorimetric snesor for GSH.

## 2.1 Chemicals and apparatus

Zinc nitrate hexahydrate ($Zn(NO_3)_2 \cdot 6H_2O$) and dimethylimidazole (2-MIM) were purchased from Sigma-Aldrich. Sodium borohydride ($NaBH_4$) were purchased from Aladdin Reagent Co., Ltd. (Shanghai, China). Glutathione (GSH) and hexachloroplatin (Ⅳ) acid were purchased from Rhawn Reagent Co., Ltd. (Shanghai, China), peroxide ($H_2O_2$, 30%) were bought from Sinopharm Chemical Reagent Co., Ltd. (Beijing, China).

Scanning electron microscope (SEM) images and energy-dispersive X-ray analysis (EDAX) are characterized by Phillips XLZ 30. Transmission electron microscope (TEM) images are characterized by JEOL JEM-F200(Japan). FT-IR spectrometer (Nicolet iS10) was used to record the FT-IR spectrum. UV–vis absorption spectrum was recorded on a UV–vis spectrophotometer (Shimadzu UV-2600, Japan). Powder X-ray diffraction (XRD) patterns were performed on an X-ray diffractometer (Rigaku D/max- 2500).

## 2.2 Synthesis of Pt@ZIF-8

The solutions of ($Zn(NO_3)_2 \cdot 6H_2O$) (0.585 g dissolved in 4 mL deionized water) and 2-methylimidazole (11.35 g dissolved in another 40 mL water) were prepared. Two solutions were rapidly mixed together under stirring at 35 °C. The produced milky solution was stirred for 30 min, and the product was separated by centrifuging. The obtained ZIF-8 was washed with deionized water for at least 3 times and dried overnight in an oven.

30 mg of ZIF-8 was dissolved in 4 mL water. 1 mL of $Na_2PtCl_6$ and 8 mg/mL of $NaBH_4$ were added into the solution. The color of the solution changes from white to brownish black. After stirring at room temperature for 30 min, centrifuge at 5000 rpm for 3 min, wash with deionized water for 3 times, and then disperse the suspension in 7 mL of deionized water and store it at 4 °C [25].

**2.3. Peroxidase-like activity assays**

The peroxidase activity of Pt@ZIF-8 was detected by TMB and $H_2O_2$ system. 0.1 mL of Pt@ZIF-8 dispersion was added into 2 mL sodium acetate buffer (pH=5), then TMB (0.1 mL, 2 mM) and $H_2O_2$ (0.1 mL, 100 mM) were added to the mixture. The mixing solution was then incubated at room temperature for 5 min. Then, the absorbance value at 652 nm was measured and the solution was photographed.

**2.3. Colorimetric detection of GSH**

First, 0.1 mL Pt@ZIF-8 dispersion, 0.1 mL TMB (5 mM), 0.1 mL $H_2O_2$ (100 mM) were added into 2.1 mL sodium acetate buffer, the solution was incubated for 35 minutes, and then GSH was added to the mixture. Afterwards, the absorbance at 652 nm was determined by a spectrophotometer. A standard curve was established by GSH concentration and the change in absorbance at 652 nm.

**3.1 Characterization of Pt@ZIF-8**

The preparation method is illustrated in Scheme 2. First, Zn ions and dimethylimidazole ligands form a three-dimensional skeleton structure of ZIF-8. Then, the Pt nanoparticles reduced by sodium borohydride were grown and adsorbed on the surface of ZIF-8, in this method we successfully synthesize Pt@ZIF-8.

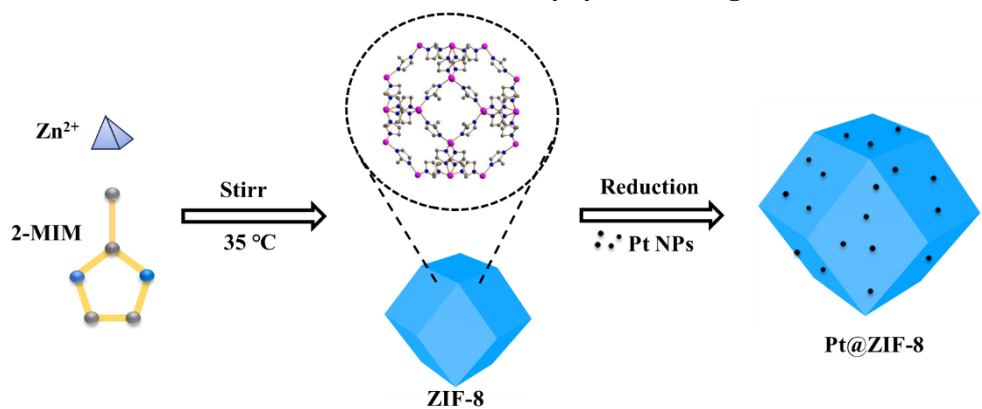

Scheme 2. Schematic illustration for preparation process of Pt@ZIF-8

In Fig. 1A and B, SEM images, the morphology of ZIF-8 was almost unchanged after a layer of Pt nanoparticles was embedded on the surface in Fig. 1B. Particle size distribution diagram shows that the particle size of Pt@ZIF-8 is slightly larger than that of ZIF-8 (Fig. S1). Pt nanoparticles are uniformly distributed on the surface of ZIF-8 (Fig. 1C, D), the element content of Pt is about 1%, and the average particle size of Pt nanoparticles is about 3 nm (Fig. 2).

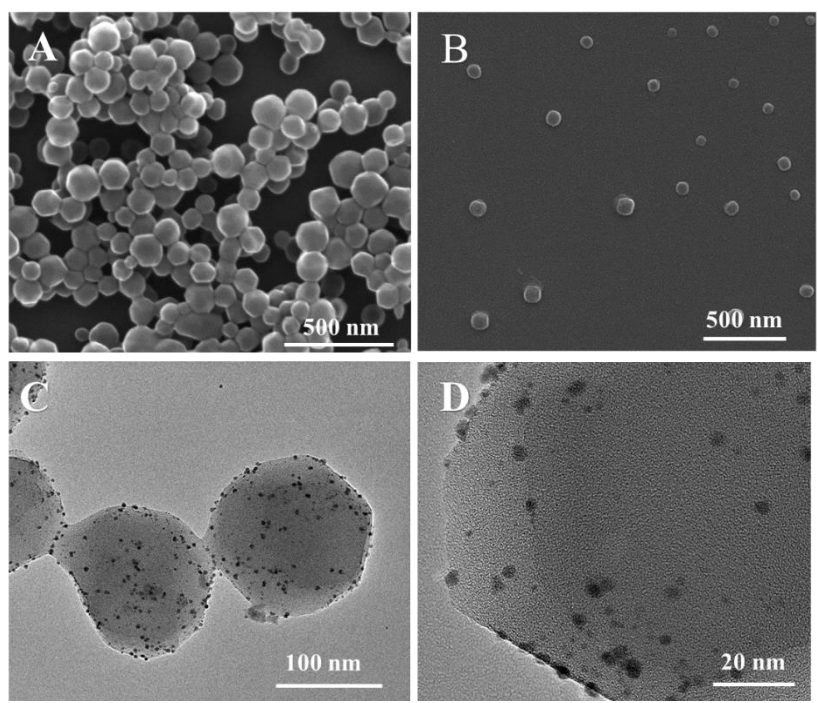

Fig. 1. (A) SEM of ZIF-8, (B) SEM of Pt@ZIF-8, (C, D) TEM of Pt@ZIF-8

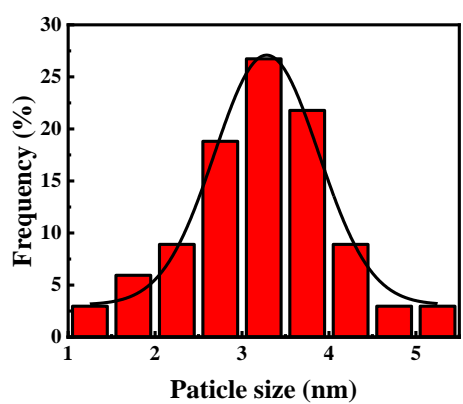

Fig. 2. Size distribution of Pt nanoparticles

The Pt@ZIF-8 nanoplates were further characterized by XRD (Fig. 3A), the main peaks of synthesized ZIF-8 were located at 7.32°, 10.34°, 12.7° and 18.02°, which were agreed with previously reported works [26]. These diffraction peaks were indexed as (001), (002), (112) and (222), respectively. Pt@ZIF-8 is almost the same as the peak position of ZIF-8, which proves that the organic skeleton material of ZIF-8 remains intact during the reaction process. There is a weak additional peak at 40°, corresponding to the (111) plane of Pt [27]. The smaller peak intensity may be due to the low content of Pt, more evidence is shown in the EDS (Fig. S1). In FT-IR spectra (Fig. 3B), The obvious peak at 953~1511 cm$^{-1}$ is related to the imidazole ring stretching of ZIF-8, and the characteristic peaks at 1580 cm$^{-1}$ and 761 cm$^{-1}$ correspond to the signals of C=N and Zn-O bonds [28-29]. Compared with ZIF-8, the strength of some peaks at Pt@ZIF-8 decreases, for example, the peak intensity at 1580 cm$^{-1}$ decreased obviously, which may be because surface attachment of Pt nanoparticles on the stretching vibration of C=N bonds.

The X-ray photoelectron spectroscopy (XPS) survey shows the Pt@ZIF-8 contained Zn, O, N, C, and Pt elements [30] (Fig. 4A). In Fig. 5B, the peaks at 1021.2 eV and 1044.2 eV correspond to Zn $2P_{3/2}$ and Zn $2p_{1/2}$ [31]. Fitting the high-resolution XPS spectrum of Pt 4f (Fig. 4C), the peaks at 73 eV and 76.33 eV correspond to $Pt^{2+}$, and the presence of +2 valence may be due to insufficient reduction of chloroplatinate by $NaBH_4$, and the peaks at 71.08 eV and 74.41 eV correspond to $Pt^0$ [32]. From the area of Pt 4f deconvoluted peaks show that the atomic percentages of $Pt^0$ and $Pt^{2+}$ are calculated to be 49.65% and 50.35%, respectively. Zeta potential diagram shows that the potential of ZIF-8 is 26.6 mV, while that of Pt@ZIF-8 is 5.09 mV (Fig. 4D). ZIF-8 has a higher positive potential, while chloroplatinate is negatively charged. Based on the Brunauer−Emmett−Teller (BET) method (Fig. 5), the surface area of Pt@ZIF-8 is calculated to be 980.4 m²·g⁻¹, which is slightly smaller than that of ZIF-8 (1,142 m²·g⁻¹). The above analysis proves the successful synthesis of Pt@ZIF-8.

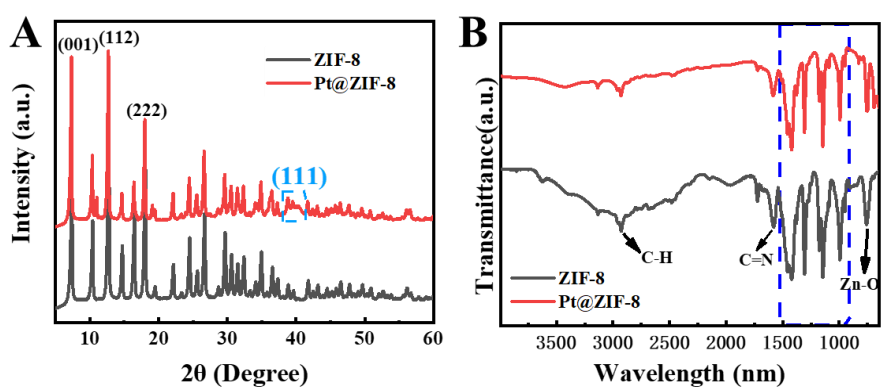

Fig. 3. (A) XRD of ZIF-8 and Pt@ZIF-8, (B) FT-IR of ZIF-8 and Pt@ZIF-8

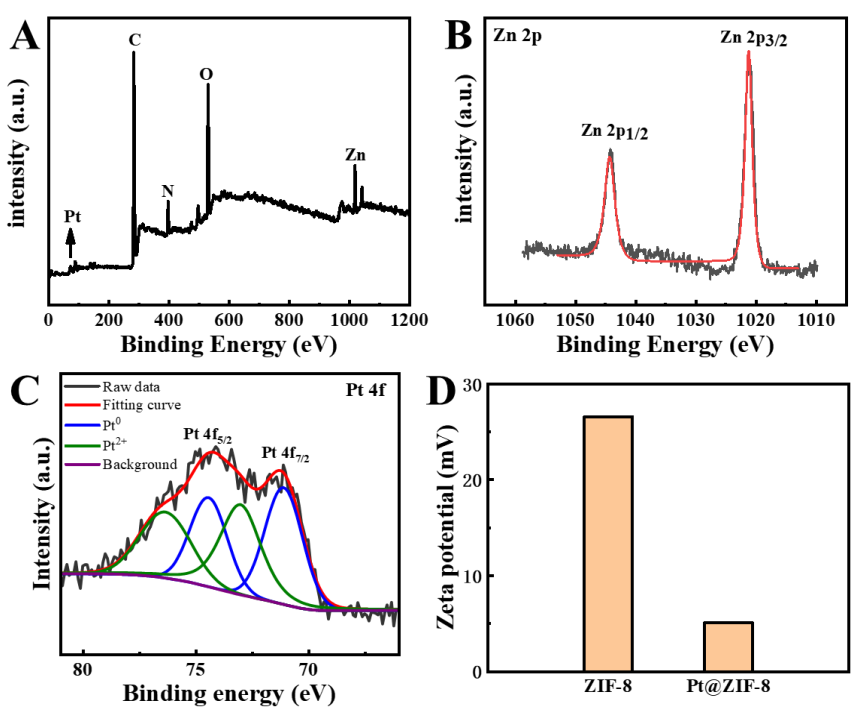

Fig. 4. (A) XPS spectra of Pt@ZIF-8, (B) HR-XPS spectra of Zn 2p, (C) Pt 4f, (D) ZIF-8 and Pt@ZIF-8 Zeta potentials

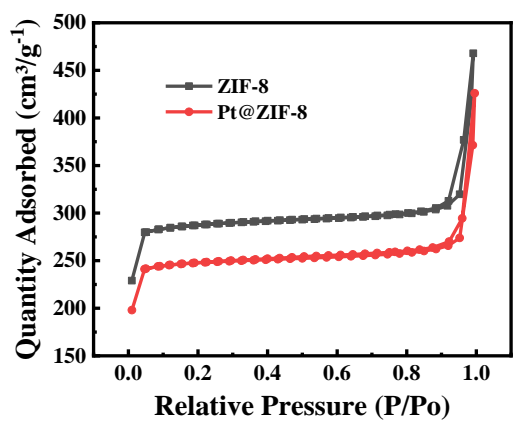

Fig. 5 Nitrogen adsorption-desorption isotherm of Pt@ZIF-8 and ZIF-8

## 3.2 Peroxidase-like activity of Pt@ZIF-8

The POD-like properties of Pt@ZIF-8 are detected by exploring the oxidation of TMB, the solution will have an absorption peak at 652nm when TMB is oxidized to oxTMB. To check the characteristic in our Pt@ZIF-8, $H_2O_2$ and TMB were employed as substrates for colorimetric measurements. As displayed in Fig. 6A, the TMB substrate is rapidly oxidized to TMBox mediated by $H_2O_2$ under the catalysis of Pt@ZIF-8, while ZIF-8 has no peroxidase properties, indicating that the main catalytic effect is the dispersion of Pt nanoparticles on the surface of ZIF-8. To further investigate whether Pt@ZIF-8 exists oxidase activity, we injected $N_2$ into the reaction system, as shown in Fig. 6B. It was found that after $N_2$ was injected, the absorbance of Pt@ZIF-8+TMB and Pt@ZIF-8+TMB+$H_2O_2$ systems decreased slightly, so Pt@ZIF-8 has slight oxidase activity. It is proved that Pt@ZIF-8 can react reactive oxygen species produced by $H_2O_2$ with TMB to produce oxTMB. In order to explore the types of free radicals decomposed by $H_2O_2$, IPA was used as the scavenger of hydroxyl free radicals. With the increase of IPA content, the absorbance at 652 nm gradually decreased until there was no absorption peak (Fig. 7).

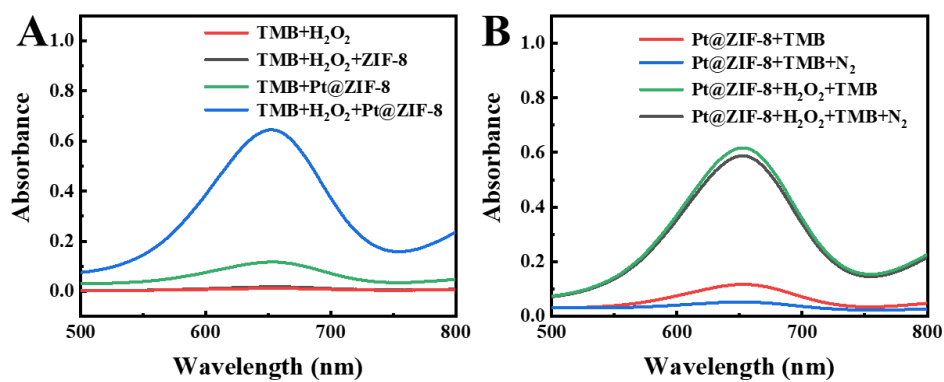

Fig. 6. (A) UV-VIS absorption spectra of different TMB systems, TMB + $H_2O_2$, TMB + Pt@ZIF-8, TMB + $H_2O_2$ + Pt@ZIF-8, TMB + ZIF-8 + $H_2O_2$, (B) UV-VIS absorption spectra of different TMB systems: TMB + Pt@ZIF-8, TMB + Pt@ZIF-8 + $N_2$, TMB + $H_2O_2$ + Pt@ZIF-8, TMB + $H_2O_2$ + Pt@ZIF-8 + $N_2$

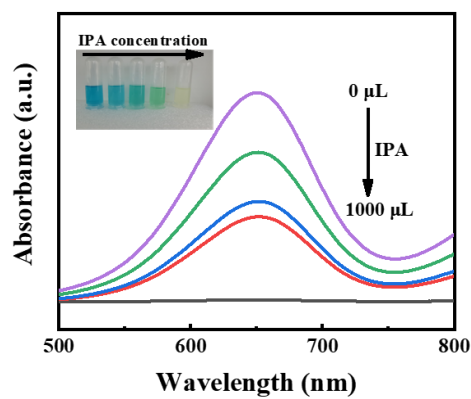

Fig. 7. UV–vis absorption spectra of oxTMB with varied volume of IPA. Inset: photograph of samples.

## 3.3 Experimental condition optimization

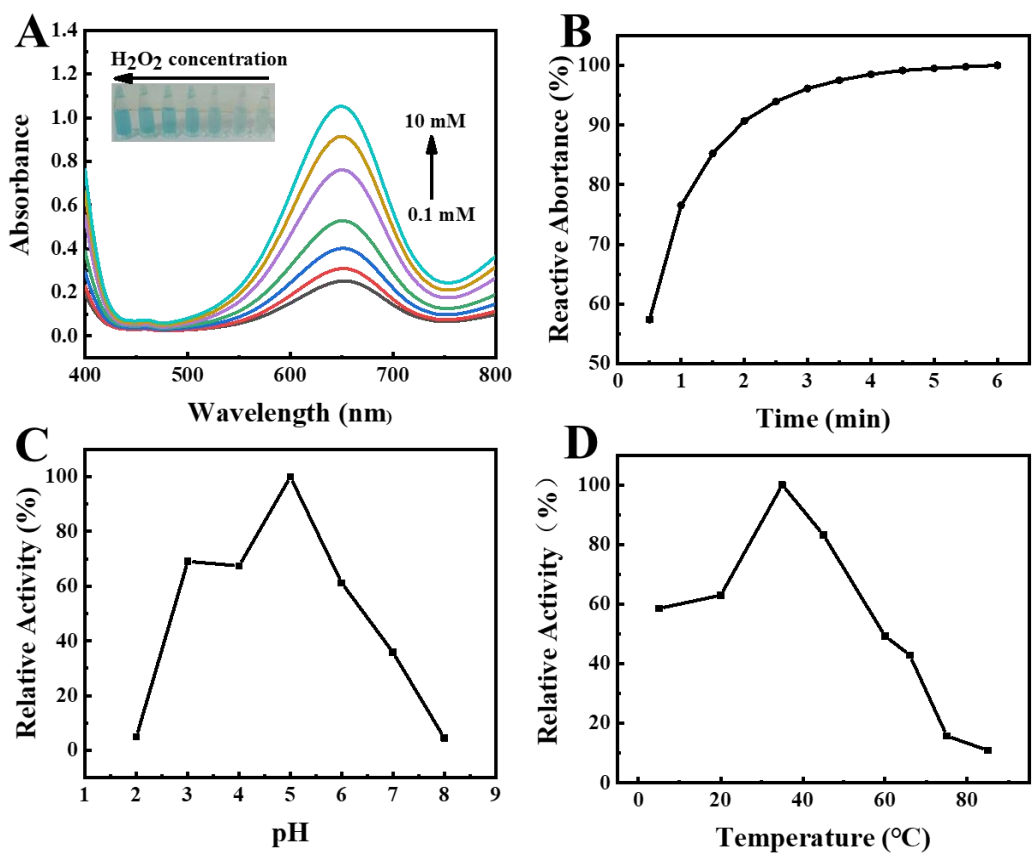

Fig. 8. (A) UV–vis absorption spectra of oxTMB with varied concentration of $H_2O_2$. Inset: photograph of samples. The catalytic activity under different conditions, (B) incubation time, (C) pH, and (D) temperature

As shown in Fig. 8A, with the $H_2O_2$ concentration increased, the absorbance of the system at 652 nm increased significantly, and it can be observed from the real picture that the color of the solution deepened. The activity of enzymes will be affected by the surrounding environment. In order to better explore the catalytic activity of nanase, we further discussed the change of the catalytic activity of Pt@ZIF-8 with reaction temperature, pH value and reaction time. With the increase of reaction time, the absorbance gradually increased, and the absorbance barely changed after 3 min. The influence of pH value on the enzyme-catalyzed reaction cannot be ignored (Fig. 8C), The enzymatic activity reached the best when the pH was 5. This experiment explored the changes of enzyme activity in the range of 5~85 °C, as shown in Fig. 8D, the optimal temperature of Pt@ZIF-8 nanase is 35 °C.

### 3.4. Kinetic measurement

According to Lambert-Bier's law A=ε·c·b. In this colorimetric experiment, the light-absorbing substance is oxTMB, where A is the dimensionless absorbance. The thickness of the colorimetric dish used in the experiment is 1 cm. The molar absorption coefficient of oxTMB is 39,000 $M^{-1}cm^{-1}$. Then calculate the catalytic reaction rate V=C/T. The enzyme kinetics curve was fitted by the Michaelis-Menten equation $V = \frac{V_{max}[S]}{(K_m+[S])}$, where $V_{max}$ is the maximum reaction rate of the enzymatic reaction, $[S]$ is the substrate concentration, and $K_m$ is the characteristic constant of the enzyme.

The Michaelis-Menten constant ($K_m$) and reaction rate ($V_{max}$) of Pt@ZIF-8 are calculated according to the Michaelis-Menten equation and listed in Table S1. The $K_m$ value of $H_2O_2$ of Pt@ZIF-8 was lower than that of Hep-Pt NCs, Pt NPs-PVP and Au NPs, indicating that Pt@ZIF-8 has a higher affinity for $H_2O_2$ than the above materials. Moreover, the reaction has a faster reaction rate because the $V_{max}$ value of Pt@ZIF-8 is higher. The $K_m$ for TMB is smaller than HRP, indicating that Pt@ZIF-8 has more affinity for TMB than natural horseradish peroxidase.

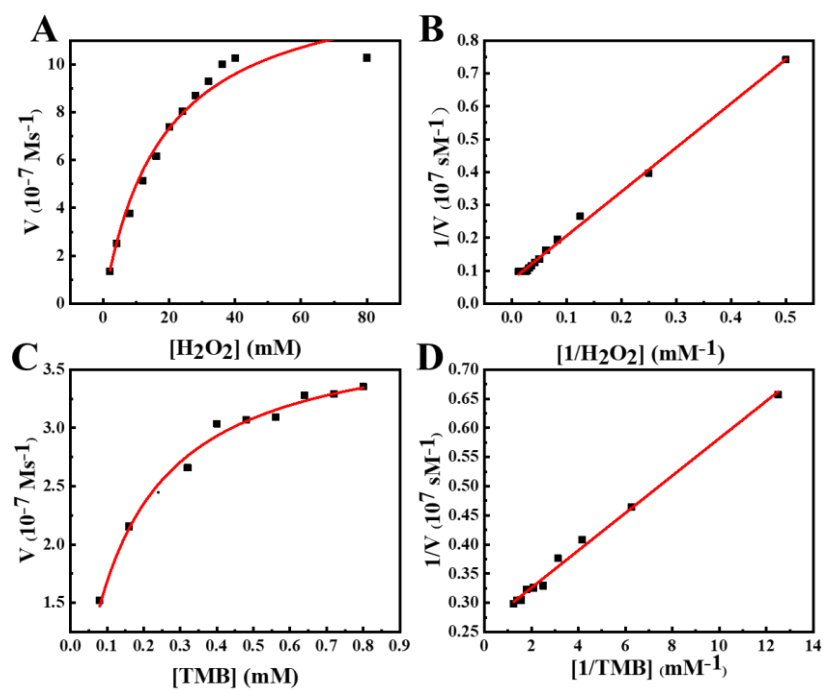

Fig. 9. Steady-state kinetic analysis. (A) and (C) reaction rate curves. (B) and (D) double reciprocal plot of (A) and (C) respectively.

### 3.5 Colorimetric detection of GSH

Reduced GSH can react with oxTMB to produce oxidized GSH and TMB, and the blue oxTMB solution becomes colorless [38], so we can monitor the content of GSH according to the color change of TMB solution. As shown in Fig. 10A, when GSH is added to TMB solution, it can be observed that the color gradually becomes lighter with the increase of GSH concentration, and the color change is monitored by UV-VIs absorption spectrum. It can be seen from the absorption spectrum that the absorbance value at 652 nm decreases with the increase of GSH concentration, and the absorbance change at 652 nm of the reaction system is calculated, $\Delta A = A_0 - A$. $A_0$ is the absorbance of TMB solution without adding GSH, and A is the absorbance of GSH solution with different concentrations. The concentration of ΔA has a linear relationship with GSH, with the linear range ranging from 10 to 1500 μM (Fig. 10B). In the low concentration range ranging from 10 to 90 μM, GSH is positively proportional to the absorbance change value, and is also positively correlated between 100 and 1500 μM, as shown in Fig. 10C and D.

The detection limit LOD=3σ/S, where σ is the linear slope and S is the standard deviation of the detected value of the blank sample. The detection limit of this experiment is 0.3185 μM, as shown in Table S2. Compared with the colorimetric analysis of GSH detected by other nanomases, Pt@ZIF-8 has a wider detection range and a lower detection limit, indicating that this reaction system has a higher sensitivity to GSH. The concentration of GSH can be quickly detected.

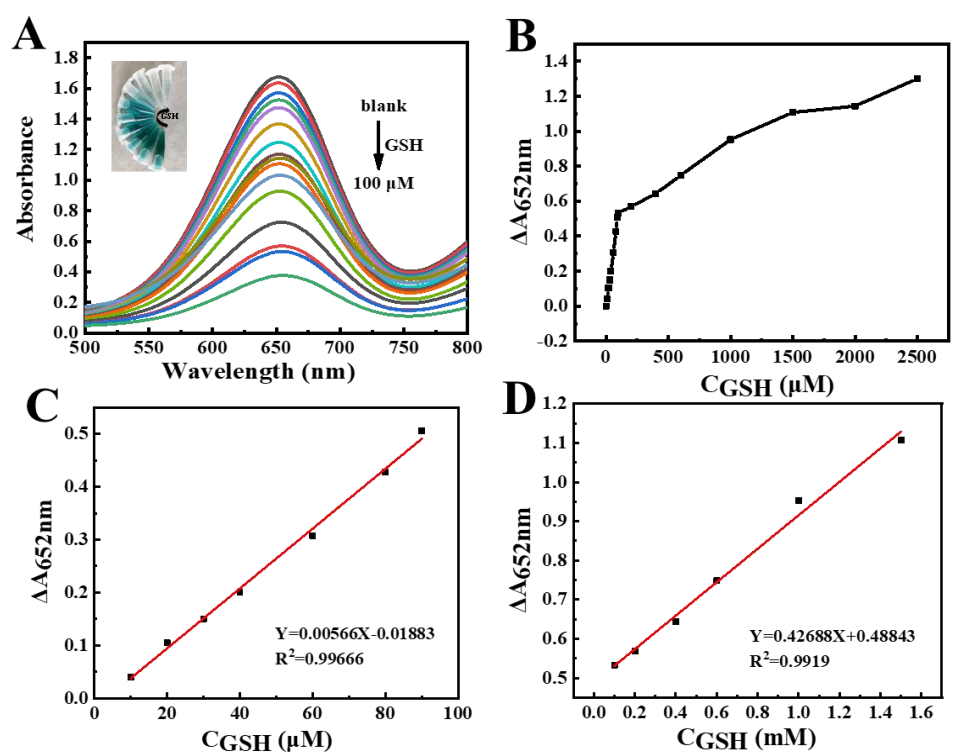

Fig. 10. (A) the change of absorbance value of TMB + $H_2O_2$ + Pt@ZIF-8 with the concentration of GSH, (B) the decrease of absorbance of the system, (C) the linear calibration diagram of GSH at low concentration, and (D) the linear calibration diagram of GSH at high concentration

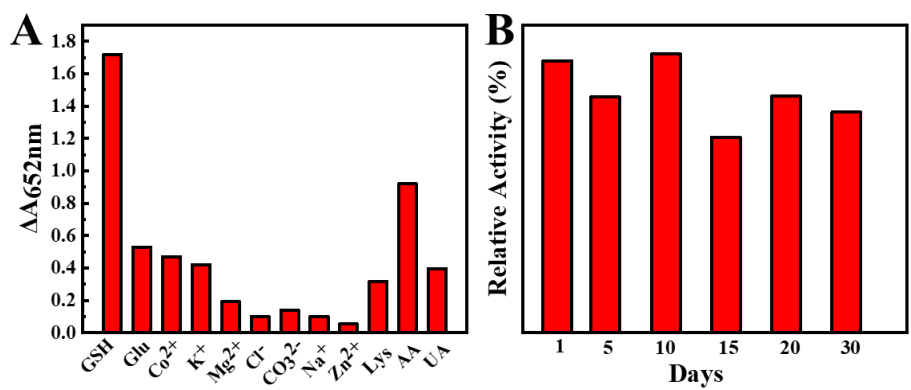

Fig. 11. (A) The $\Delta A_{652}$ responses of Pt@ZIF-8+ TMB + $H_2O_2$ system towards GSH and interferents, (B) stability test of Pt@ZIF-8

Human blood contains not only GSH, but also inorganic salts and organic molecules. Therefore, anti-interference experiments need to be conducted for the detection of GSH. As shown in Fig. 11A, when glucose, $Ca^{2+}$, $K^+$, $Mg^{2+}$, $Cl^-$, $CO_3^{2-}$, $Na^+$, $Zn^{2+}$, L-serine and urea are added, which are five times the concentration of GSH. The change of absorbance was relatively low, only when GSH was added, the absorbance decreased significantly, and when ascorbic acid was added with the same concentration, the absorbance also changed to a certain extent. However, since the concentration of reduced GSH in blood was much higher than that of AA, the influence of AA was also negligible, the experiment had good selectivity for GSH.

We tested the catalytic performance of the material within 30 days (Fig. 11B). The absorbance of the TMB system after the addition of Pt@ZIF-8 was tested on day 5/10/15/20/30, respectively. It was found that Pt@ZIF-8 had good catalytic activity within 30 days, so the material had excellent stability and could be stored for a long time.

## 4. Conclusion

In summary, ZIF-8 was synthesized by hydrothermal method, and the Pt nanoparticles were in-situ grown on the surface of ZIF-8, and the formed Pt@ZIF-8 had good peroxidase activity. Pt@ZIF-8 can accelerate the transfer rate of electrons from TMB to free radicals. Based on the good selectivity and stability of Pt@ZIF-8, the colorimetric detection of GSH was established. the detection range is from 10 to 1500 μM, the detection limit of this experiment is 0.3185 μM, which provided a further basis for their development as natural enzyme mimics and candidate substances for GSH clinical diagnosis.

**References**

[1] Zhou Y, Liu B, Yang R, et al. Filling in the Gaps between Nanozymes and Enzymes: Challenges and Opportunities. Bioconjug. Chem. 2017, 28, 2903–2909.

[2] Wu S, Tatarchuk B J, Adamczyk A J. Ethylene oxidation on unpromoted silver catalysts: Reaction pathway and selectivity analysis using DFT calculations. Surf. Sci. 2021, 708, 121834.

[3] Wu Y F, Darland D C, Zhao J X. Nanozymes—Hitting the Biosensing "Target". Sensors. 2021, 21(15), 5201.

[4] Park Y, Gupta P K, Tran V K, et al. PVP-stabilized PtRu Nanozymes with Peroxidase-like Activity and Its Application for Colorimetric and Fluorometric Glucose Detection[J]. Colloids and Surfaces B: Biointerfaces, 2021, 204, 111783.

[5] Yadav N, Jaiswal A K, Dey K K, et al. Trimetallic Au/Pt/Ag based nanofluid for enhanced antibacterial response[J]. Materials Chemistry and Physics, 2018, 218, 10-17.

[6] Wang C Y, Kang P C, Ou Y S, Li G L, et al. Room-Temperature Hydrogen Adsorption via Spillover in Pt Nanoparticle-Decorated UiO-66 Nanoparticles: Implications for Hydrogen Storage[J]. 2021, 4, 10, 11269–11280.

[7] Sharma N K, Shukla S, Sajal V. Surface plasmon resonance based fiber optic sensor using an additional layer of platinum: A theoretical study[J]. Optik, 2017, 133, 43-50.

[8] Wang H, Zhao J, Liu C, et al. Pt Nanoparticles Confined by Zirconium Metal–Organic Frameworks with Enhanced Enzyme-like Activity for Glucose Detection[J]. ACS Omega, 2021, 6, 7, 4807–4815.

[9] Li M, Li D Y, Li Z Y, et al. A visual peroxidase mimicking aptasensor based on Pt nanoparticles-loaded on iron metal organic gel for fumonisin B1 analysis in corn meal[J]. Biosensors & Bioelectronics: The International Journal for the Professional Involved with Research, Technology and Applications of Biosensors and Related Devices, 2022, 209.

[10] Khoris I M, Kenta T, Ganganboina A B, et al. Pt-embodiment ZIF-67-derived nanocage as enhanced immunoassay for infectious virus detection[J]. Biosensors and Bioelectronics 2022, 215.

[11] Hong C Y, Chen L L, Wu C Y, et al. Green synthesis of Au@WSe$_2$ hybrid nanostructures with theenhanced peroxidase-like activity for sensitive colorimetric detection of glucose[J]. Nano Research, 2022, 15, 1587–1592.

[12] Xue Y, Li H Y, Wu T, et al. Pt deposited on sea urchin-like CuCo$_2$O$_4$ nanowires: Preparation, the excellent peroxidase-like activity and the colorimetric detection of sulfide ions[J]. Journal of Environmental Chemical Engineering, 2022, 10, 2, 107228.

[13] Zhang F, Wei Y, Wu X, et al. Hollow zeolitic imidazolate framework nanospheres as highly efficient cooperative catalysts for [3+3] cycloaddition reactions[J]. J Am Chem Soc, 2014, 136(40), 13963-6.

[14] Yong P, Jia C, Bei L, et al. Separation of methane/ethylene gas mixtures efficiently by using ZIF-67/water-ethylene glycol slurry[J]. Fluid Phase Equilibria, 2016, 414, 14-22.

[15] A J W, B M Z, A Y Z, et al. Synergetic integration of catalase and Fe$_3$O$_4$ magnetic nanoparticles with metal organic framework for colorimetric detection of phenol[J]. Environmental Research, 2021, 206, 112580.

[16] Ma X B, Sui H Y, Yu Q, et al. Silica Capsules Templated from Metal−Organic Frameworks for Enzyme Immobilization and Catalysis[J]. Langmuir, 2021, 37, 3166−3172.

[17] Li X, Zheng L, Wang Y, et al. A novel electrocatalyst with high sensitivity in detecting glutathione reduced by 2-hydroxypropyl-β-cyclodextrin enveloped 10-methylphenothiazine[J]. RSC Advances, 2015, 5, 88, 71749-71755.

[18] Yu Z, Tang Y, Xing L, et al. A highly sensitive upconverting phosphors-based off–on probe for the detection of glutathione[J]. Sensors & Actuators B Chemical, 2013, 185, 363-369.

[19] Li P, Ge M, Yang L, et al. Metal coordination-functionalized Au–Ag bimetal SERS nanoprobe for sensitive detection of glutathione[J]. The Analyst, 2019, 144(2), 421-425.

[20] Muppidathi Marieeswaran, Perumal Panneerselvam. Transition metal coordination frameworks as artificial nanozymes for dopamine detection via peroxidase-like activity[J]. Materials Advances, 2021, 2, 7024.

[21] Qin R X, Feng Y H, Ding D D, et al. Fe-Coordinated Carbon Nanozyme Dots as Peroxidase-Like Nanozymes and Magnetic Resonance Imaging Contrast Agents[J]. ACS Appl. Bio Mater, 2021, 4, 5520−5528.

[22] Liu W D, Chu L, Zhang C H, et al. Hemin-assisted synthesis of peroxidase-like Fe-N-C nanozymes for detection of ascorbic acid-generating bio-enzymes[J]. Chemical Engineering Journal, 2021, 415, 128876.

[23] Song C, Liu H, Zhang L, et al. FeS nanoparticles embedded in 2D carbon nanosheets as novel nanozymes with peroxidase-like activity for colorimetric and fluorescence assay of $H_2O_2$ and antioxidant capacity[J]. Sensors & Actuators: B. Chemical, 2022, 353, 1311312021.

[24] A N B, C A K B, B J H, et al. Sensitive biosensing of organophosphate pesticides using enzyme mimics of magnetic ZIF-8[J]. Spectrochimica Acta Part A: Molecular and Biomolecular Spectroscopy, 2019, 209, 118-125.

[25] Li Y, Li S J, Bao M, et al. Pd Nanoclusters Confined in ZIF-8 Matrixes for Fluorescent Detection of Glucose and Cholesterol[J]. ACS Appl. Nano Mater. 2021, 4, 9, 9132–9142.

[26] Song D, Wang Y, Ma R, et al. Structural modulation of heterometallic metal-organic framework via a facile metal-ion-assisted surface etching and structural transformation[J]. Journal of Molecular Liquids, 2021, 334, 116073.

[27] Wang P Y, Liu J, Liu C F, et al. Electrochemical Synthesis and Catalytic Properties of Encapsulated Metal Clusters within Zeolitic Imidazolate Frameworks[J]. Chemistry-A European Journal, 2016, 22, 46, 16613-16620.

[28] Jia S R, Zhang X K, Yuan F, et al. Colorimetric Test Paper for $H_2O_2$ Determination Based on Peroxidase-Like Activity of an AuFe/ZIF-8-Graphene Composite[J]. Chemistry Select, 2022, 7, 43.

[29] Ostad M I, Shahrak M N, Galli F. Photocatalytic carbon dioxide reduction to methanol catalyzed by ZnO, Pt, Au, and Cu nanoparticles decorated zeolitic imidazolate framework-8[J]. Journal of $CO_2$ Utilization, 2020, 43, 101373.

[30] Zhou H, Zhang J, Zhang J, et al. High-capacity room-temperature hydrogen storage of zeolitic imidazolate framework/graphene oxide promoted by platinum metal catalyst[J]. International Journal of Hydrogen Energy, 2015, 40, 36, 12275-12285.

[31] Lian Z, Lu, C Q, Zhu J Q, et al. Mo@ZIF-8 nanozyme preparation and its antibacterial property evaluation[J]. Frontiers in Chemistry, 2022, 10, 1093073.

[32] Chen L Z, Klemeyer L, Ruan M B, et al. Structural Analysis and Intrinsic Enzyme Mimicking Activities of Ligand-Free PtAg Nanoalloys[J]. Small, 2023.

[33] Xi X X, Wang J H, Wang Y Z, et al. Preparation of Au/Pt/$Ti_3C_2Cl_2$ nanoflakes with self-reducing method for colorimetric detection of glutathione and intracellular sensing of hydrogen peroxide[J]. Carbon, 2022, 197, 476–484.

[34] Gao L, Zhuang J, Nie L, Zhang J, et al. Intrinsic peroxidase-like activity of ferromagnetic nanoparticles, Nature Nanotech. 2007,2, 577–583.

[35] Gu H, Huang Q, Zhang J, et al. Heparin as a bifunctional biotemplate for Pt nanocluster with exclusively peroxidase mimicking activity at near-neutral pH[J]. Colloids and Surfaces A Physicochemical and Engineering Aspects, 2020, 606, 125455.

[36] Zhang D Y, Liu H, Younis M R, et al. Ultrasmall platinum nanozymes as broad-spectrum antioxidants for theranostic application in acute kidney injury[J]. Chemical Engineering Journal, 2021, 409, 127371.

[37] X. Chen, X. Tian, B. Su, Z. Huang, X. Chen, M. Oyama, Au nanoparticles on citrate-functionalized graphene nanosheets with a high peroxidase-like performance[J]. Dal-ton Trans, 2014, 43, 7449–7454.

[38] Pt and $ZnFe_2O_4$ Nanoparticles Immobilized on Carbon for the Detection of Glutathione[J]. ACS Applied Nano Materials, 2021, 4, 9, 9479-9488.

[39] Cao X Y, Yang H, Wei Q L, et al. Fast colorimetric sensing of $H_2O_2$ and glutathione based on Pt deposited on NiCo layered double hydroxide with double peroxidase-/oxidase-like activity[J]. Inorganic Chemistry Communications, 2021, 123, 108331.

[40] Wang Y, Liu X, Wang M, et al. Facile synthesis of CDs@ZIF-8 nanocomposites as excellent peroxidase mimics for colorimetric detection of $H_2O_2$ and glutathione[J]. Sensors and Actuators B Chemical, 2020, 329, 129115.

[41] Xi J, Zhu C, Wang Y, et al. $Mn_3O_4$ microspheres as an oxidase mimic for rapid detection of glutathione[J]. RSC Advances, 2019, 9, 29, 16509-16514.

Table S1. Comparison of enzymatic kinetic parameters of Pt@ZIF-8 and other nano-enzymes

| Materials | $K_m$/mM | | $V_{max}$ ($10^{-8}$ M·s$^{-1}$) | | Ref |
|---|---|---|---|---|---|
| | $H_2O_2$ | TMB | $H_2O_2$ | TMB | |
| HRP | 3.70 | 0.41 | 8.71 | 10.0 | [34] |
| Hep-Pt NCs | 165 | 0.016 | 11.98 | 7.18 | [35] |
| Pt NPs-PVP | 79 | 2.3 | 1.72 | 2.81 | [36] |
| Au NPs | 45.83 | 0.74 | 10.69 | 12.15 | [37] |
| Pt@ZIF-8 | 18.55 | 0.122 | 138.33 | 38.1 | This work |

Table S2 Performance contrast of various nanomaterials in detecting GSH by colorimetry technique

| Materials | Linear range(μM) | LOD(μM) | Ref |
|---|---|---|---|
| Pt/NiCo-LDH NCs | 50000-500000 | 3770 | [39] |
| Au/Pt/Ti$_3$C$_2$Cl$_2$ | 50-10000 | 10.2 | [33] |
| CDs@ZIF-8 | 0-100 | 1.04 | [40] |
| Mn$_3$O$_4$ microspheres | 5.0-60 | 0.889 | [41] |
| Pt@ZIF-8 | 10-1500 | 0.3185 | This work |

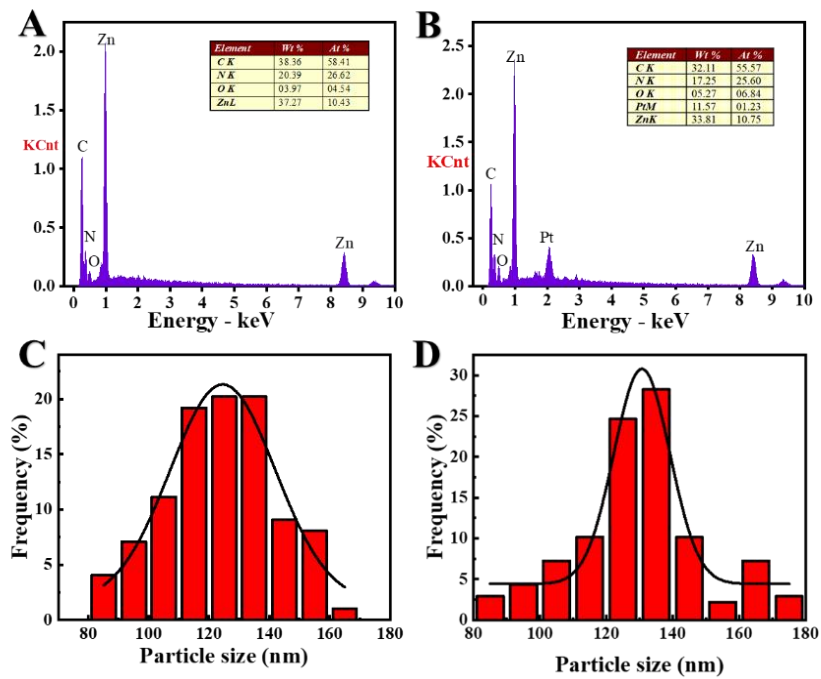

Fig. S1. (A) EDS of ZIF-8, (B) EDS of Pt@ZIF-8, (C) ZIF-8 grain size distribution, (D) Pt@ZIF-8 grain size distribution